# 3D Single-pixel imaging with active sampling patterns and learning based reconstruction

Xinyue Ma,1,2,†  and Chenxing Wang,1,2,†,*

[1]School of Automation, Southeast University, 2# Sipailou, Xuanwu, Nanjing, 210096, China

[2] Key laboratory of Measurement and Control of Complex Systems of Engineering, Ministry of Education, Southeast University, Nanjing, 210096, China

†Authors contributed equally to this work.

*cxwang@seu.edu.cn

## Abstract：

Single-pixel imaging (SPI) is significant for applications constrained by transmission bandwidth or lighting band, where 3D SPI can be further realized through capturing signals carrying depth. Sampling strategy and reconstruction algorithm are the key issues of SPI. Traditionally, random patterns are often adopted for sampling, but this blindly passive strategy requires a high sampling rate, and even so, it is difficult to develop a reconstruction algorithm that can maintain higher accuracy and robustness. In this paper, an active strategy is proposed to perform sampling with targeted scanning by designed patterns, from which the spatial information can be easily reordered well. Then, deep learning methods are introduced further to achieve 3D reconstruction, and the ability of deep learning to reconstruct desired information under low sampling rates are analyzed. Abundant experiments verify that our method improves the precision of SPI even if the sampling rate is very low, which has the potential to be extended flexibly in similar systems according to practical needs.

## 1. Introduction

Single-pixel imaging (SPI) is one of the computational imaging methods [1-4], which has attracted much attention due to its excellent capabilities for imaging applications with limited data transmission bandwidth or non-visible light imaging [5-7]. A SPI system often works by projecting a series of sampling patterns to a target scene first, then capturing the light intensity of the scene successively using a single-pixel detector (SPD), and finally reconstructing the target scene from the captured data sequence. If the captured signal carries depth information, 3D SPI can be achieved. An early 3D SPI technique is implemented based on the shape-from-shading scheme [8], i.e., a series of random binary patterns are projected to an object, then the light intensity from four directions is captured by four SPDs and four images are restored correspondingly, finally the 3D shape is reconstructed with the images, based on the photometric stereo [9]. This method requires highly optimized deployment of the SPDs. The time-of-flight (TOF) based method is another form of 3D SPI [10], which estimates the depth according to the flight time of pulses. Thus, a pulse source with high frequency and high intensity is required for practical use. Inspired by the fringe projection profilometry (FPP), a Fourier-based 3D SPI method [11] is proposed. And some new illuminating methods are proposed to modulate and acquire depth information to realize 3D SPI [12, 13].

In the above methods, sampling strategy and reconstruction algorithm are the most important factors for SPI. The sampling strategy usually refers that some patterns are projected to an object by a projector, a DMD, or other illumination devices. The commonly used patterns are random patterns [14,15], including random grey-scale patterns and random binary patterns, and the latter ones are more often adopted because they can be generated and projected much faster [16]. However, the random patterns perform the sampling blindly and so some spatial points may be missed to be sampled if the sampling rate is low, resulting in poor imaging quality. Increasing the sampling rate helps solving this problem, but this leads to a large time cost for acquisition and computation. A Hadamard matrix composed of 1 and -1 is employed to perform the sampling in complementary form [17], but this method requires double SPDs or double binary patterns. The above-mentioned Fourier basis can also be taken as a type of sampling pattern [11], but it also needs multiple times of sampling patterns and thus greatly increases the imaging time.

As for the reconstruction algorithm, various algorithms are proposed based on mathematical theories in the early stage. A total variation based image compressive sensing recovery is proposed using the non-local regularization [18]. The orthogonal matching pursuit (OMP) [19] is an often-used one for SPI. In each iteration, the OMP selects the variables which have the highest correlation with the current residual, and finally solves these variables to restore the signal. However, these algorithms usually require a high sampling rate and complex calculation, but even so, the imaging quality is still unsatisfactory. Recently, deep learning methods have been introduced for reconstruction, which perform better compared to traditional methods [20, 21]. Deep learning methods highly depend on the dataset which is required to be large and diversiform. However, creating a dataset by real experiment is time-consuming and laborious. Some methods simulate the sampling process to 2D images to obtain the dataset and train a deep convolutional network [20] or generative adversarial network (GAN) [21] to realize 2D SPI. However, these simulation methods cannot be used for preparing the dataset of 3D SPI. Due to this difficulty, the application of deep learning for 3D SPI is relatively rare.

In response to the above problems, this paper proposes a 3D SPI scheme with an active sampling strategy and learning-based reconstruction. With a projector being the illumination device, we design some windows as the sampling patterns, which will be projected to the target object in sliding form. The shape, size, location, and the sliding order of the windows can all be designed flexibly according to the needs, which implies the active sampling. The light intensity is captured by an SPD, in front of which a grating sheet is placed. Therefore, the captured intensity values can be reordered to form a low-resolution binary fringe image. Then the depth map is reconstructed from this fringe image by deep neural networks where two approaches of constructing the networks, end-to-end and two-stage, are compared. Our training data are generated flexibly by a virtual system based on graphics to increase the performance of the networks. Abundant experiments illustrate the effect of our method even if the sampling rate is very low.

## 2. The proposed 3D SPI method

## 2.1 The proposed 3D SPI scheme

The proposed 3D SPI scheme is based on the FPP technique [11,12] and combined with deep learning. Figure 1 shows the structure of the system, which includes a projector, a single-pixel detector (SPD), a grating sheet, a digital acquisition card, and a computer. A series of sampling patterns are projected onto the object, and the SPD collects the signals successively. Finally, the depth map is reconstructed from the collected signals. The proposed method has a similar procedure to other SPI systems. Moreover, it has the novelty below:

(1) Our sampling strategy is active and controllable, and the sampling patterns can be designed flexibly according to the needs.
(2) We can easily restore a binary fringe image by reordering the captured values, which provides the spatial information additionally. This makes up for the shortcoming that the SPD lacks the spatial resolution capability and simplifies the complex optimization of SPI to the problem of super resolution.
(3) We introduce deep neural networks as the reconstruction method that improve the quality of reconstruction, and we also introduce the graphic technique to create a virtual system which can generate a large amount of training data needed for 3D SPI.

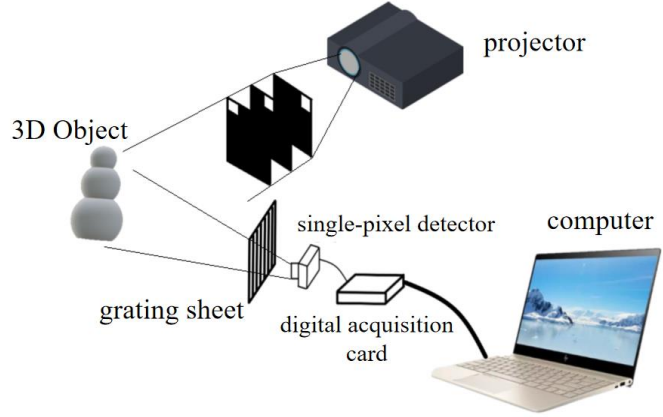

Fig. 1. The structure of the 3D SPI system in this paper.

## 2.2 The designed sampling patterns

The design of the sampling patterns is inspired by the image down-sampling. For an image $S$, each pixel can be extracted by

$$s_{ij} = \sum\sum E_1 \cdot S, (1 \leq i \leq W, 1 \leq j \leq H) \tag{1}$$

where $s_{ij}$ is the pixel at position $(i, j)$ of $S$; W and H denote the width and height of $S$, respectively; $E_1$ is a matrix with the same size as $S$, which has value 1 at $(i, j)$ and 0 at other positions, and the subscript 1 means that the area size of the window is 1.

For a SPI system, the $E_1$ matrix is used as a sampling pattern and projected to the target scene, then an illumination intensity is captured by the SPD. Consequently, a complete image $S$ can be obtained after W×H times of projection and detection by making the window, value 1 at $(i, j)$, move and traverse all positions of the target scene. Figure 2 displays the images of $E_1$ patterns.

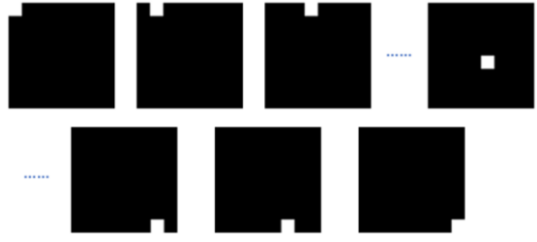

Fig. 2. The $E_1$ matrices (W×H) that can traverse all positions of a target scene.

To improve the imaging speed, the size of the window can be increased, and then the sampled pixel can be replaced as

$$s'_{ij} = \text{Round}[(\frac{1}{N}\sum\sum E_N \cdot S)], (1 \leq i \leq W, 1 \leq j \leq H) \qquad (2)$$

where Round[ • ] means a rounding operation, and $E_N$ is a matrix whose window has $N$ pixels. Thus, Eq. (2) performs like that the image S is averagely down-sampled with sampling rate $1/N$.

With this type of sampling pattern projected, the signal captured by the SPD is the total light intensity of the area illuminated by the window, which is equivalent to implement the term $\sum\sum(E_N \cdot S)$ in Eq. (2). Therefore, each detected signal can be viewed as an integrated value in a local region, thus a low-resolution image can be reorder after the target scene is scanned by the window. Furthermore, since there is a grating sheet placing in front of the SPD, the formed image is thus a low-resolution fringe image. As the example in Fig. 3, an original high-resolution fringe image is shown as Fig. 3(a), and Fig. 3(b) is the low-resolution fringe image obtained by sampling Fig. 3(a) with the above sampling method, where the window size is $N=4$.

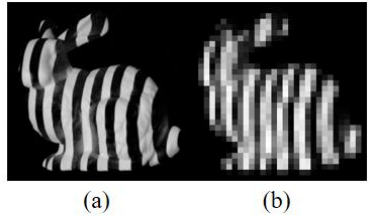

(a)            (b)

Fig. 3. (a) A fringe image; (b) the low-resolution fringe image restored after sampled.

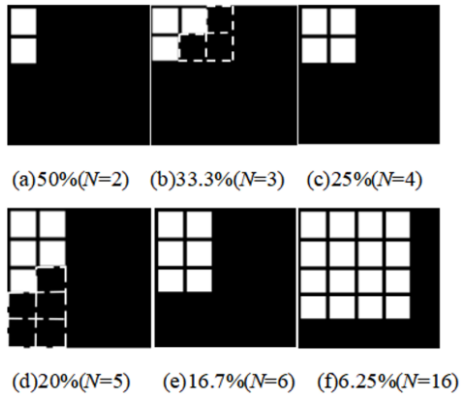

(a)50%(N=2)   (b)33.3%(N=3)   (c)25%(N=4)

(d)20%(N=5)   (e)16.7%(N=6)   (f)6.25%(N=16)

Fig. 4. The window with different $N$ (different sampling rate).

With different value of $N$, the shape of the window can be designed as in Fig. 4. For an area sampled by an irregular window, the window in the next pattern can be inverted as marked by the dotted line in Fig. 4(b) and Fig. 4(d), so that the windows are projected pairwise to cover a regular shape. In addition, the long edge of the window is recommended to be consistent with the fringe orientation to ensure that the details of the fringe deformation can be sampled sufficiently.

The usual strategy of window sliding can choose the way of raster scanning, from the left to right and top to down. For different requirements, the sliding strategy can be changed in different order, for example, the sliding scanning can be a swirl shape starting from the center of a target object. All these reflect the flexibility of our method.

2.3 The reconstruction method based on deep learning networks

2.3.1 Two reconstruction approaches with deep learning networks

The neural network is used to reconstruct the depth map from an input low-resolution fringe image. To achieve this, we introduce two approaches: end-to-end and two-stage.

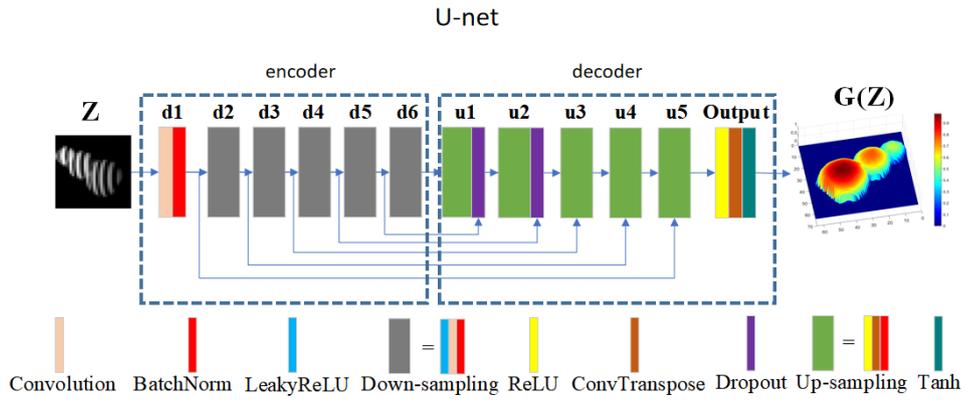

Fig. 5. The structure of U-Net.

For the end-to-end approach, the well-known U-net [22] is used to establish the mapping of the low-resolution fringe image to the depth map. The structure of the U-net is shown in Fig. 5, which consists of an encoder and a decoder. The encoder consists of six down-sampling blocks to deeply extract features. The decoder consists of six up-sampling blocks to restore the expected output. And there are skip connections between the encoder and decoder. The sufficient up-sampling layers can decode the features to obtain a high-resolution output [22], while the skip connections transmit the features of different scales to different up-sampling layers to ensure that both global information and local details can be retained better for final results.

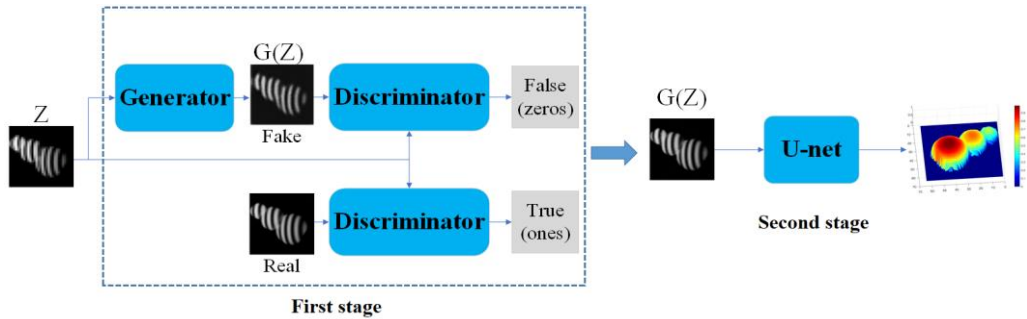

Fig. 6. The principle of two-stage network.

For the two-stage approach, a high-resolution fringe image is reconstructed by a generative adversarial network (GAN) first, and then the depth map is retrieved by another network.

Considering the better performance of image generating, we choose a conditional GAN, pix2pix [23], for the first stage. The principle of pix2pix is illustrated in Fig. 6. As common cGAN does, the pix2pix generates an image through an adversarial learning procedure. The generator adopts the U-net as the basis since it can extract the underlying features of the input. The discriminator adopts patchGAN, which implements the discrimination in patches divided from the input. More details of pix2pix can be found in [23]. For the second stage, another U-net is also applied to restore the depth, like the end-to-end approach above.

In these two approaches, the loss functions are also designed and chosen with careful considerations. The U-net is adopted for its powerful ability of feature extraction, and the features of structure similarity are emphatic in our method, thus the Structure SIMilarity (SSIM) function is introduced to design the loss of U-net, that is

$$L_{U-net} = 1 - \text{SSIM}(U(Z), real\_depth), \quad (3)$$

where $Z$ is the input low-resolution fringe image, $U(Z)$ is the output depth map, $real\_depth$ is the ground truth, and SSIM( · ) is defined as

$$\text{SSIM}(u,v) = \frac{(2\mu_u \mu_v + c_1)(2\sigma_{uv} + c_2)}{(\mu_u^2 + \mu_v^2 + c_1)(\sigma_u^2 + \sigma_v^2 + c_2)}, \quad (4)$$

where $\mu_u$ is the mean value of output $u$, $\mu_v$ is the mean value of ground truth $v$, $\sigma_u^2$ and $\sigma_v^2$ are the variance of $u$ and $v$, respectively, $\sigma_{uv}$ is the covariance of $u$ and $v$, $c_1$ and $c_2$ are constants to avoid dividing by zero.

The pix2pix has more loss functions because of its more complex structures. For the generator taking U-net as the basis, the loss is designed as

$$L_G = \text{MSE}(D(Z, G(Z)) - ones) + 100 \times (1 - \text{SSIM}(G(Z), real\_fringe)), \quad (5)$$

where D( · ) and G( · ) denote the discriminator and generator, respectively, $Z$ is the input and $G(Z)$ is the output from the generator, MSE( · ) is to calculate the mean square error, $real\_fringe$ is the ground truth, $ones$ is a matrix with all values being 1, and SSIM( · ) is also used to restrain the structural information. For the discriminator, the commonly used MSE loss is adopted because it is sufficient to evaluate whether the $G(Z)$ is close to the real one.

2.3.2 Some issues for implementing the approaches

*I. The generation of dataset*

For the end-to-end approach, we need to prepare low-resolution fringe images and the corresponding depth maps as the input and output, respectively. For the two-stage approach, the high-resolution fringe images are needed additionally as the output of the first stage.

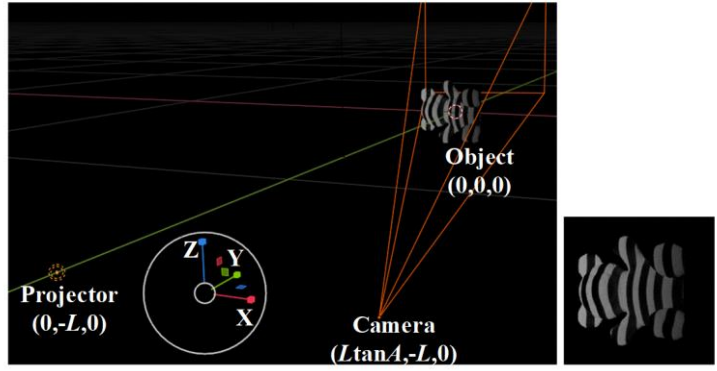

Fig. 7. The virtual system to generate fringe images.

To flexibly and conveniently generate the dataset, we adopt a graphic software, Blender, to build a virtual imaging system, as shown in Fig. 7. There is a virtual camera, and a virtual projector which is realized by the built-in virtual point light source. The parameter $L$ represents the distance from the projector to the object, and $A$ is the angle between the projector and the camera. Some detailed settings can refer to [24]. With this virtual system, we generate the dataset including:

a) Depth maps, generated by 3D models of objects chosen from the Thingi10K [25];

b) Low-resolution fringe images, generated by projecting **binarized** fringes to objects first, capturing the **binarized** fringe images and down-sampling them as described in section 2.2;

c) High-resolution fringe images, generated by projecting **sinusoidal** fringes to objects first and capturing the images.

*II. Some factors to let the simulated data close to the real ones*

The virtual system generates data in an ideal environment, but the actual system is interfered by some factors unavoidably, which should be analyzed and added to the virtual system to let the simulated data closer to reality. The virtual imaging system contains a camera and a projector. Therefore, the factors influencing the system are nothing more than the environment illumination and the location deployment of hardware.

Since the SPD is interfered with ambient light easily, the experiments in this paper are conducted in the dark. However, these disturbances cannot be avoided completely in practice and appear as random noise. Therefore, we add some random noise with intensity ranging in [0.04, 0.14] to the images normalized to [0, 1].

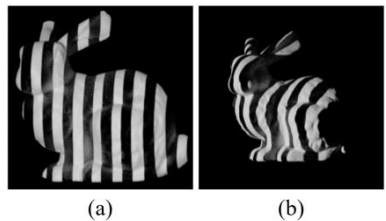

(a)        (b)

Fig. 8. The fringe images captured by different angles of camera and projector: (a) 5°, (b) 40°.

The depth change of a surface can be extracted from the fringe deformation in a fringe image by the principle of triangulation. The angle between the camera and projector needs to be considered carefully. The fringe will hardly deform if the angle is too small, while some information will be lost if the angle is too large, as shown in Fig. 8. In this paper, the angle is set at about 15° referring to [12]. In the virtual system, the angle is set to change within the range [13°, 17°] randomly in the data generation, to be adapted to the uncertainties of actual systems.

*III. The relationship between the grating period and the window size*

In the actual system, we form the binarized fringes by placing a grating sheet in front of the SPD, while we sample the target scene by the designed patterns. According to the Nyquist law, a constraint below should be satisfied

$$M \leq \frac{T}{2}, \tag{6}$$

where $T$ is the fringe period and $M$ denotes the window size. Unlike the variable $N$ in section 2.2, $M$ represents the number of pixels in the horizontal direction. An example of sampling an ideal binarized fringe image is shown in Fig. 9. If $M$ is small enough as required in Eq. (6), the sampling rate is high, and more details can be retained, as shown in Fig. 9(a). If Eq. (6) is not satisfied but it is still limited that $M \leq T$, Fig. 9(b) shows that the fringe can be sampled with some distortion. Once $M$ is set larger than $T$, the expected information is impossible to be restored due to the aliasing of under-sampling.

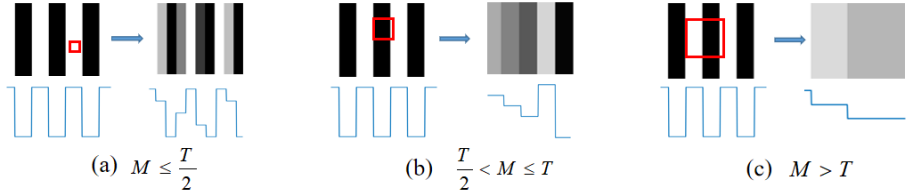

(a) $M \leq \frac{T}{2}$    (b) $\frac{T}{2} < M \leq T$    (c) $M > T$

Fig. 9. The different cases of sampling with: (a) $M \leq \frac{T}{2}$; (b) $\frac{T}{2} < M \leq T$; (c) $M > T$.

For the case in Fig. 9(b), correctly restoring the information is difficult for traditional methods, because they need to rigorously follow the Eq. (6) to cope with the theoretical derivation. But it is easier for our approaches, since there are a large amount of training data providing prior information, and the model by deep supervised learning can predict some basic structural information from these priors. Figure 10 visualizes the procedure results for the samples in Figs. 9(a) and 9(b) respectively, where d2 and u5 are two layers of the generator (U-net) of pix2pix. Even if the sampling rate has been as low as $M > T/2$, some effective features still can be extracted, ensuring that the final results can be decoded correctly.

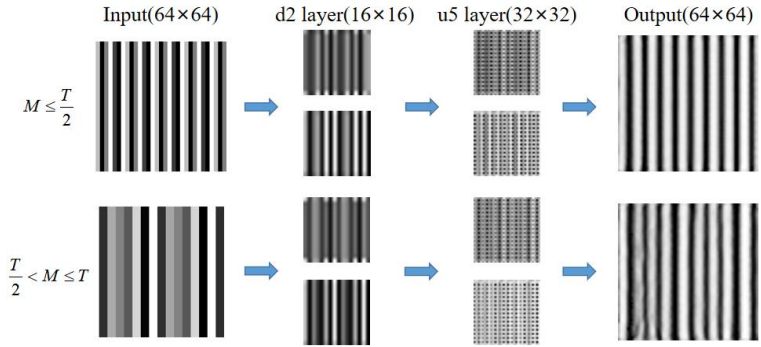

Fig. 10. Visualization of reconstruction results corresponding to Fig. 9(a) and 9(b).

In the real system, the period of binarized fringes, *T* in Eq. (6), is determined after a grating sheet is deployed in an actual system. The window size, *M* in Eq. (6), should be considered according to the limits of *M* analyzed above. For actual implementation, we can test the limit size of *M* to set a working range roughly and then set *M* within the range according to specific imaging requirements, i.e., $M \leq T/2$ if higher accuracy is required, and $T/2 < M \leq T$ if faster speed or lower sampling rate is desired. To ensure the generalization of the network, the dataset is prepared with *T* and *M* being diverse and consistent with actual use.

## 3. Simulation and experiment

3.1 Parameter settings

The test objects are taken from the 3D models in the Thingi10K dataset. 624 objects are selected and divided into a training set and a test set, with a ratio of 0.85: 0.15. In the virtual system, each object is rotated 48 times around the X, Y, and Z axes respectively, so there are 76,464 sets of data for training and 13,392 sets of data for test. The size of each image is 64×64 pix$^2$.

There are three kinds of sampling rates in the training dataset, namely 50%, 25%, and 6.25%, which are set with a ratio of 1:1:2. The data with a low sampling rate is set with a larger number to enhance the learning ability for this type of data. In addition, the fringe period is set in a range to enhance the generalization ability of the network, which is randomly set between 6-8 pixels.

The hyperparameters to train the U-net or pix2pix are the same, i.e., the learning rate is 0.0001, the batch size is 8, the dropout ratio is 0.5, and the LeakyRelu layer parameter is 0.2, etc. The ADAM optimizer is used, and all the networks are trained with 20 epochs. The entire network construction and training process are completed under the PyTorch framework. The computer is configured with Intel Core i7-10700 CPU, 32GB RAM, and GeForce RTX 3090 graphic card.

3.2 Simulations

*I. The comparison of our two approaches*

We compare the end-to-end and two-stage approaches with the 13392 sets of test data. The results are quantified in Table 1 where *α* is the mean error, *δ* is the mean square error (MSE), and *γ* is the largest absolute error. The two-stage approach reconstructs a high-resolution fringe image first, which can provide richer information for reconstructing the depth and so shows better than the end-to-end, so we adopt the two-stage approach for the following comparisons only.

Table 1 Comparison of the two strategies.

| Method | Sampling rate | | | | | | | | |
|---|---|---|---|---|---|---|---|---|---|
| | 50% | | | 25% | | | 6.25% | | |
| | *α* | *δ* | *γ* | *α* | *δ* | *γ* | *α* | *δ* | *γ* |
| End-to-end | -0.0031 | 0.0066 | 0.9427 | -0.0032 | 0.0069 | 0.9467 | -0.0040 | 0.0090 | **0.9607** |
| Two-stage | **-0.0011** | **0.0063** | **0.9407** | **-0.0011** | **0.0066** | **0.9448** | **-0.0020** | **0.0087** | 0.9618 |

*II. The comparison of our sampling patterns to the commonly used random patterns*

The proposed sampling pattern is also compared with another commonly used one, the random pattern. Our two-stage approach is used for both of the two types of sampling patterns. We refer to [8] to design the random patterns. We conduct simulations with dataset sampled by two sampling patterns respectively. The quantified results on the test set are listed in Table 2, and we also display some results in Fig. 11 and Fig. 12, where the positions with obvious differences are marked with white boxes. Obviously, the result of our patterns show much better than the ones of random patterns.

Table 2 Comparison of the two sampling patterns both using our two-stage approach.

| Sampling method | Sampling rate | | | | | | | | |
|---|---|---|---|---|---|---|---|---|---|
| | 50% | | | 25% | | | 6.25% | | |
| | $\alpha$ | $\delta$ | $\gamma$ | $\alpha$ | $\delta$ | $\gamma$ | $\alpha$ | $\delta$ | $\gamma$ |
| Random patterns | -0.0146 | 0.0117 | 0.9674 | -0.0099 | 0.0103 | 0.9745 | -0.0045 | 0.0151 | 0.9852 |
| Our patterns | **-0.0011** | **0.0063** | **0.9407** | **-0.0011** | **0.0066** | **0.9448** | **-0.0020** | **0.0087** | **0.9618** |

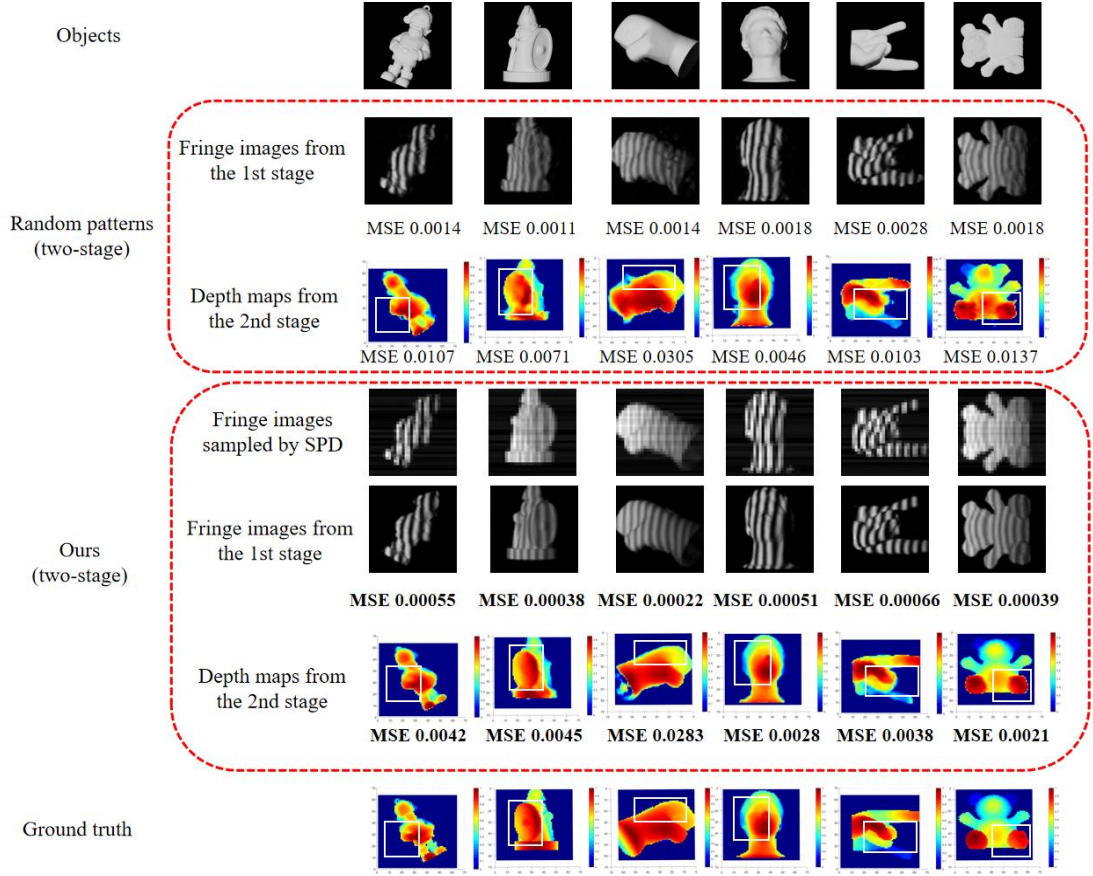

Fig. 11. Depth maps obtained for several objects (sampling rate: 25%).

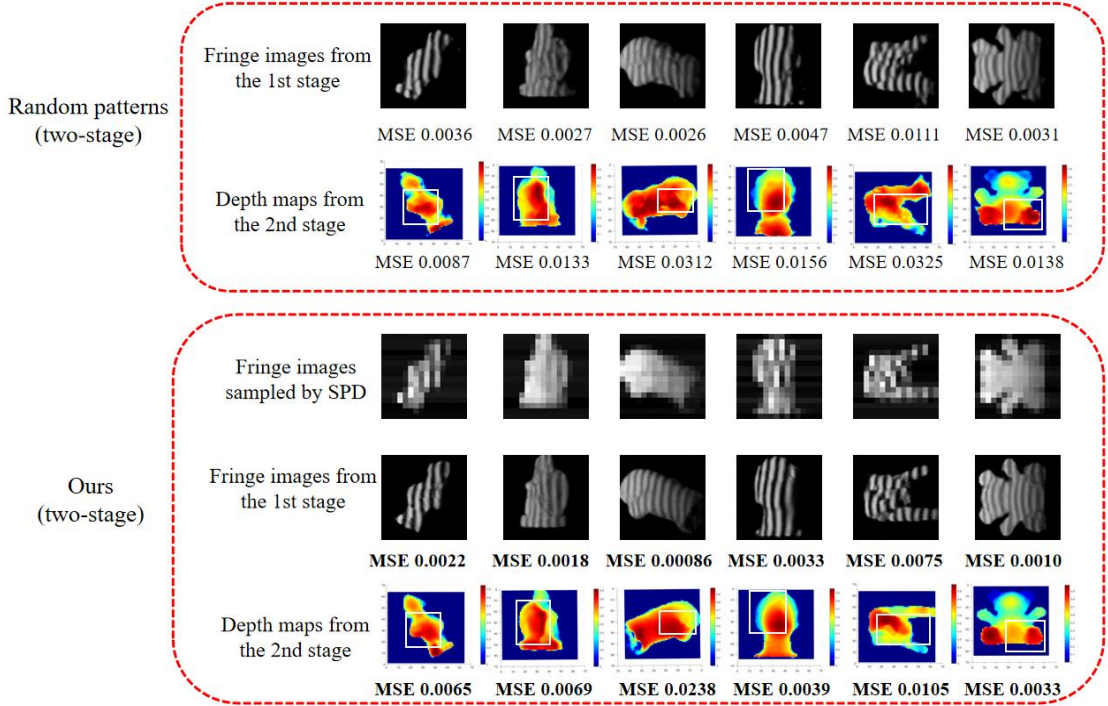

Fig. 12. Depth maps obtained for several objects (sampling rate: 6.25%).

*III. The comparison of different methods*

In the above experiment, our two-stage network is used to compare the two different sampling patterns. We also compare the results wholly by our method with the method in [12, 26]. The test data and the test model are all from the opened source in [26]. Figures 13(a)-13(c) are the results of our method when the sampling rate is 50%, 25%, and 6.25%, respectively. And Fig. 13(d) is the depth map of the method in [12] with a sampling rate of 30%. Compared to the ground truth, our results show much better than the one of [12] even if the sampling rate is very low.

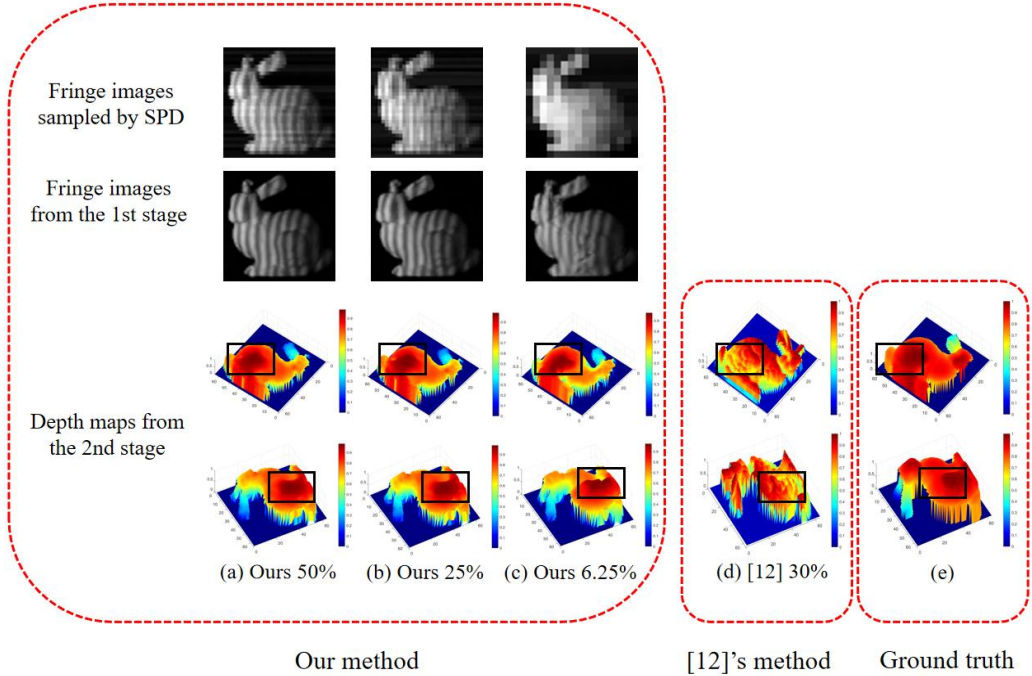

Fig. 13. By our method with sampling rate: (a-c) 50%, 25% and 6.25% respectively; (d) by method in [13] with sampling rate 30%; (e) ground truth.

3.3 Real experiment

The experimental system is shown in Fig. 14, where the projector is DLP4500, the SPD is PDA100A2, and the data acquisition card is NI USB-6216. The grating sheet is customized by high-precision 3D printing of which the space width is 4mm.

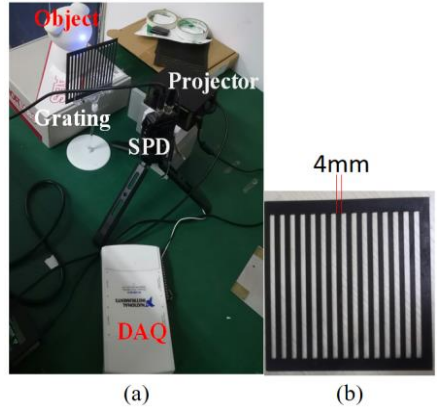

Fig. 14. (a) Experimental system; (b) The grating sheet.

First, we test the limit size of $M$ roughly. A light source is placed at the location of the SPD to form the binarized fringes, then $M$ is adjusted to be $T/2$ or $T$ visually according to the projected window in the fringes, as shown in Fig. 15. Then the working range of $M$ can be recorded. In this experiment, $T \approx 6$ (pixel), so $M$ can be set within [1, 6] (pixel). In this experiment, in order to obtain a qualified result, we set the largest $M$ as 4.

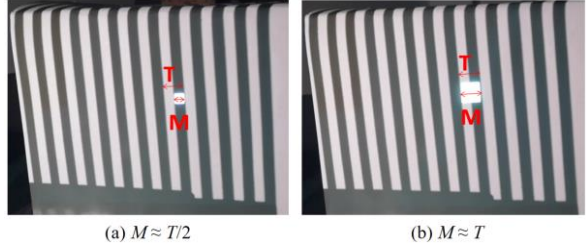

Fig. 15. The limit size of $M$ adjusted by visually letting: (a) $M \approx T/2$; (b) $M \approx T$.

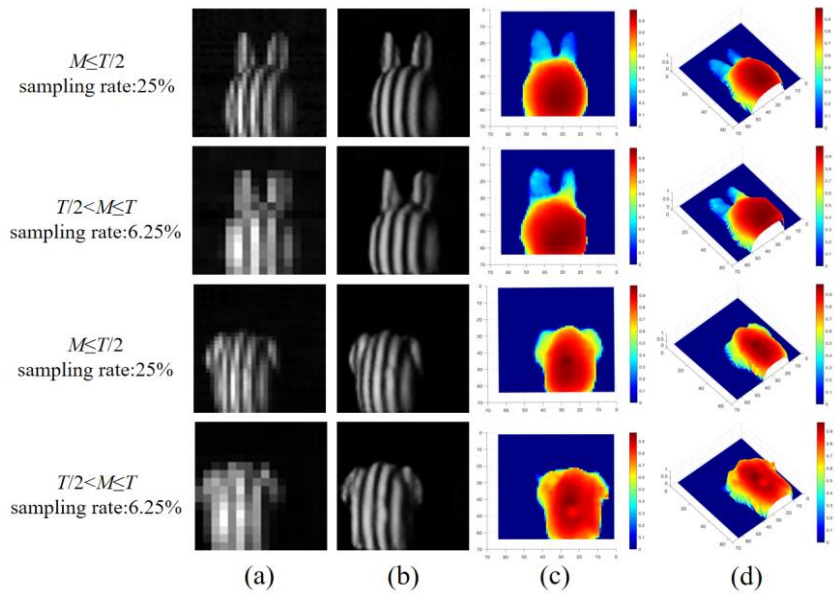

Fig. 16. (a) The low-resolution fringe image collected by the single-pixel detector; (b) the high-resolution fringe image output by our network; (c)-(d) the resulted depth maps.

Two objects are tested to verify our method. The test results are shown in Fig. 16. It is obvious that both shapes and details are captured and reconstructed better when $M \leq T/2$. And we can quickly capture information and reconstruct a rough shape at a very low sampling rate when $M > T/2$.

We also compare our patterns with random patterns. The sampling rate is 25%, and our two-stage approach is implemented for both two. Figure 17 shows that our patterns can extract and reconstruct better edge information presented in white boxes.

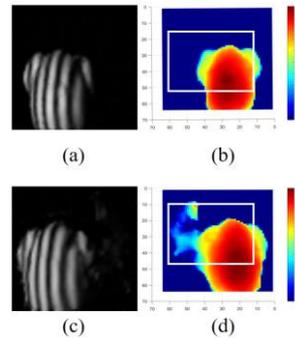

Fig. 17. The fringe image and the depth map reconstructed by our two-stage approach using:
(a-b) our sampling patterns; (c-d) the random patterns.

## 4. Conclusion

In this paper, a practical 3D SPI scheme is proposed. An active strategy of sampling is proposed, which ensures a complete sampling in SPI. Then the deep learning methods are introduced for reconstructing the depth map, where the end-to-end approach and two-stage approach are compared. The probabilities of successfully reconstructing the data even if it is sampled in a low rate is analyzed deeply in this paper. To solve the data problem for deep learning, a graphic software is introduced to build a virtual system to generate diverse data samples. Abundant experiments prove the superiority of the proposed method even if the sampling rate is very low. Furthermore, the active sampling patterns and scanning order can be changed flexibly according to the needs, implying the potentiality of our method being extended in practice.

## Funding

## Declaration of Competing Interest

The authors declare that they have no known competing financial interests or personal relationships that could have appeared to influence the work reported in this paper.